\documentclass{ws-procs9x6}
\newcommand{\ltsima} {$\; \buildrel < \over \sim \;$}  
\newcommand{\gtsima} {$\; \buildrel > \over \sim \;$}  
\newcommand{\lta} {\lower.5ex\hbox{\ltsima}}  
\newcommand{\gta} {\lower.5ex\hbox{\gtsima}}  

\begin{document}

\title{New insights on unification of radio-loud AGN}

\author{Marco Chiaberge}

\address{Istituto di Radioastronomia - CNR \\
via P. Gobetti 101 \\ 
40129 Bologna, Italy \\ 
E-mail: chiab@ira.cnr.it}

\maketitle

\abstracts{The  radio-loud  AGN unification  model associates powerful  
radio  galaxies with radio-loud quasars  and blazars.  In analogy with
the radio-quiet  scheme, the nuclear  regions of objects  showing only
narrow emission lines  in their  optical  spectrum are thought  to  be
obscured  to our line-of-sight by  a geometrically and optically thick
dusty  "torus".  In objects showing  broad emission  lines we directly
observe the innermost parsecs around the central  black hole, i.e. the
broad line region and the accretion disk.  Radiation  from the base of
the relativistic jet dominates the  overall emission of blazars,  that
are seen almost pole-on.  Although the broad  picture seems to be well
established, there are several  fundamental aspects that are  still to
be understood.    HST studies  have recently shed    new light on many
issues, from the properties of the nuclei to the structure of the host
galaxies.}

\section{The standard unification picture}

The  standard  picture  for  unification   of   both radio-quiet   and
radio-loud AGN  is based  on  the anisotropy  of the  nuclear emission:
objects intrinsically  identical appear different to observers located
along different  viewing directions.  In addition  to the  presence of
1-to-100pc-scale   absorbing tori, radio-loud    AGN have two powerful
relativistic jets, which constitute a   further source of  anisotropy.
The jets  emerge from the innermost regions,  and  propagate along the
rotation axis  of the accretion disk formed  around the  central black
hole\cite{urrypad,barthel}.  The torus blocks  the direct view of both
the accretion disk and   the surrounding broad  line region  (BLR)  to
observers  located perpendicularly  to  the  jet axis.   The  extended
(kpc-scale)  narrow emission line region  (NLR)  is visible to such an
observer,  which would classify  that object  as  a narrow-line  radio
galaxy (NLRG).  If  the line-of-sight forms a very  small angle to the
jet axis, the emission is  dominated by non-thermal radiation from the
jet,  which is  strongly  boosted by relativistic effects\footnote{The
observed  integrated   flux  is  enhanced  by   a   factor $\delta^4$,
$\delta=[\Gamma(1-\beta\cos\theta)]^{-1}$   is  the  beaming   factor,
$\theta$ is viewing  angle to the jet  axis, and $\Gamma$  and $\beta$
are the  bulk Lorentz factor and velocity  of the jet, respectively.}.
In that case, such object would  appear to us  as a {\it blazar}, i.e.
either  a flat spectrum radio quasar  (FSRQ) or a BL~Lac, depending on
whether strong emission lines  (EW $ > 5$  \AA) are present or absent,
respectively.  For intermediate  viewing angles ($\theta > 1/\Gamma$),
where the jet radiation appears  less boosted or even de-boosted, both
the accretion disk and the BLR  become visible, and the object appears
to us as  a steep spectrum radio quasar  (SSRQ) or, for lower  nuclear
luminosities, a broad line radio galaxy (BLRG).

Among   the various observations that   can be performed  to prove the
validity of such a unification scenario, the detection of a hidden BLR
seen through  scattered  radiation is  considered as  one of  the most
solid and  direct tests\cite{ski85,ski90}.  In fact,  scattering mirrors
provide a ``periscope'' for viewing the nucleus from other directions

\subsection{Open problems}

The ``zeroth-order'' unification   scheme  described  above  generally
accounts for the observational properties  of radio-loud AGN. However,
several crucial issues are still open, and need further investigation.
First of all, there is clear evidence that such a scheme must be split
into two different models, in order to account  for unification of low
and high power objects separately.  At high luminosity, radio galaxies
with ``edge-brightened'' (FR~II) morphology  are unified  with quasars
(SSRQ  and   FSRQ),  while for    lower powers, radio    galaxies with
``edge-darkened'' morphology  (FR~I)  are  believed to  be  the parent
population  of BL~Lac  objects.   The need  for two  different schemes
comes from the  ``absence'' of strong emission lines  in both  BL Lacs
and  FR~I, while  strong,    high   excitation  lines are  a    common
characteristic of powerful FR~II  and  quasars.  Furthermore, in   the
low-power  scheme,  the ``intermediate''  class analogous  to BLRG and
SSRQ appears   to  be  missing,  since   only very  few   examples  of
broad-lined FR~Is are   known  (see  section \ref{fr1}).    Thus,  the
nuclear structure of  low and high  power objects  may differ  in some
crucial aspect.

Let me now summarize what are, in my  opinion, the most important open
problems.  i) The lack of significant broad emission lines in FR~I and
BL~Lacs implies that, in the  framework of the unification model, {\it
there is no need for the presence of obscuring tori in these sources}.
However, molecular tori  are thought to be present  in all  other AGN,
and they  are also believed to  function as a  fuel  reservoir for the
active nucleus; ii) environment:    FR~Is and BL~Lacs seem   to prefer
different environments.  FR~I  are predominantly found in high density
clusters\cite{zirbel97},     while BL~Lacs     seem    to  avoid  rich
clusters\cite{owen}; iii)  a few  BL~Lacs  show  faint (and  variable)
broad  lines\cite{corbett};  iv) a fraction   of the BL~Lac population
shows a radio morphology more typical of FR~II\cite{kollgaard92}; v) a
substantial  fraction     of     narrow-lined  FR~II   have   atypical
low-ionization optical  spectra ($L_{[OIII]}<  L_{[OII]}$)  similar to
those  of FR~I,   while   quasars have $L_{[OIII]}>   L_{[OII]}$;  vi)
although polarized broad  emission lines have  been observed  in a few
NLRG (e.g.  3C~234), it is still unclear what is the incidence of such
a phenomenon  among radio-loud AGN.   Furthermore, in many objects the
origin    for polarization is  not     clear: scattering and  dichroic
extinction are  two  possible interpretations\cite{cohen99}; vii)  the
quasar fraction in  complete  samples of  radio  galaxies is  strongly
dependent on luminosity\cite{willott}, but it  is unclear whether this
happens because the inner radius of the torus increases for increasing
luminosity of the central quasar, or because of the rise of a distinct
population of ``starved'' quasars at low luminosities.

These   ``phenomenological''  issues   translate  into   more  general
questions on  the physical properties  of radio-loud AGN: are FR~I and
FR~II (or  BL~Lacs  and  quasars)  physically different?    Are  their
central BH masses different?  What   are the properties of   accretion
around the BH? Are their jets different? Do all RL  AGN have a BLR? In
the  following I will    try and summarize  the current  observational
scenario.

\section{Jets in radio galaxies and quasars}

Relativistic jets are seen emerging from the very innermost regions of
the   active  nucleus, on    sub-pc scales\cite{junor99,giovannini01}.
Radio observations show  that jets are  relativistic in both  FR~I and
FR~II radiogalaxies on   parsec scales. Superluminal motions of  radio
components are observed in  a large number   of quasars and, although
less prominently, also in BL~Lacs.  In low power radio galaxies, while
proper motions of radio   components are commonly observed on   small
scales, the jet slows  down  to sub-relativistic  velocities
over distances of $\sim 1-10$ kpc\cite{laing99}.  Superluminal motions of
optical components  have also  been   observed with  the HST in  M~87,
having  apparent speed  in the  range  $4-6c$ thus strongly supporting
unification with BL~Lacs\cite{sbm99}.

Recently, significant step forward  in our knowledge of the  structure
of jets  has  been taken.  In particular,   high resolution radio data
have shown the presence of  transverse structures in jets associated
with both  low  and high power radio  galaxies,  and even on  the VLBI
scale\cite{giovannini}.    Their  ``limb-brightened''  appearance   is
currently  explained as due to  the presence  of velocity structures in
the jet: a fast, highly relativistic ``spine'' ($\Gamma \sim 15$) and a
slower external layer which moves  at a slower speed, possibly because
of the interaction with the surrounding ambient medium.  In FR~II such
structures appear to persist on  very large scales  ($>> 10 kpc$, e.g.
3C~353\cite{swain}), indicating  that at least  the ``spine'' of FR~II
jets does not suffer  substantial deceleration.   Independent evidence
for  large-scale high-speed jets in quasars   and FR~II is provided by
X-ray  Chandra    observations.   X-ray  emission   from   large-scale
extragalactic jets is best interpreted as  a result of inverse Compton
emission from relativistic   particles scattering off seed  photons of
the cosmic microwave background\cite{tavecchio,celotti01}.  Such model
requires the bulk Lorentz factor of the jet (spine) to be $\sim 15$.

\section{Host galaxies}

A  fundamental prescription of the  unification models is that unified
classes must    share the  same properties   as   far as the  extended
(unbeamed) characteristics are concerned. Therefore, the properties of
the host galaxy    are  crucial parameters  for   testing  unification
scenarios.  Quasar hosts spanning  a range of redshift  $z=0.1-2$ have
been extensively studied with  the  HST\cite{dunlop}.  Radio loud  QSO
are hosted   by bright  ($L>L^{*}$)   massive ellipticals,  which  are
similar in magnitude and morphology to radio galaxies hosts.  The same
result holds for BL~Lacs host galaxies, which appears to be similar to
FR~I   hosts,   in  substantial   agreement    with  the   unification
scenario\cite{urry00,scarpa00}.   However, the  absence of  BL~Lacs in
rich cluster environments\cite{owen}  and the  lack  of dust lanes  in
BL~Lac hosts,  compared  with those of  FR~Is in  which dust lanes are
ubiquitous, are still issues to be addressed.

\section{Black hole masses}

The correlations between the mass  of the central  black hole and some
fundamental parameters of the host galaxy  (either the central stellar
velocity  dispersion  $\sigma$ or the  optical magnitude  of the bulge
$L_{\rm   bulge}$)\cite{tremaine,mclure}   are considered  as powerful
tools to  estimate ${\rm M}_{\rm BH}$ within  an accuracy  of a factor
2--3.  But the  determination  of the  velocity dispersion  and/or the
optical   magnitude   of  the  host  is   not  always straightforward,
especially if a  strong nuclear component is present,  such  as in the
case of quasars and BL~Lacs. BH masses of  BL~Lacs have been estimated
by\cite{barth,falomo}  using the  ${\rm M}_{\rm  BH}-\sigma$ relation,
obtaining values   in  the range $5\times  10^{7}-   1\times 10^9 {\rm
M}_\odot$.  The distributions  of  central BH  masses in  BL~Lacs  and
radio galaxies    (of  all species)  appear   to   be consistent, thus
supporting unification.  However,   to the best   of my knowledge,   a
detailed comparison of well defined and statistically complete samples
has not been performed yet.

Concerning high power objects, the  BH mass of a  sample of radio loud
QSO  and radio galaxies  has been estimated using  HST images of their
host     galaxies  and    the   ${\rm    M}_{\rm   BH}-L_{\rm  bulge}$
relation\cite{mclure}.  These objects share a  common range in $M_{\rm
BH}$ but, interestingly, it appears that radio loud AGN have BH masses
confined to M$_{\rm BH}> 10^{9} {\rm M}_\odot$. However, other authors
do not find any  relationship between radio-loudness and central black
hole mass\cite{woo}.

\section{The HST view of FR~I and FR~II nuclei: implications for
unification}

\label{fr1}

\begin{figure}[ht]
\epsfxsize=10.0cm   
\epsfbox{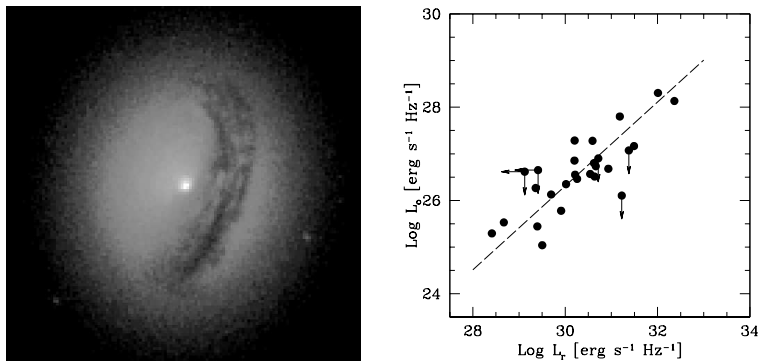}
%\hspace{1.5cm}
%\epsfxsize=4.5cm   
%\epsfbox{chiaberge_fig1b.ps}
\caption{{\bf Left}: the central 6 arcsec of 3C~449 as seen with HST/WFPC2. {\bf
Right}: optical CCC luminosity  versus  radio core luminosity for  3CR
FR~Is.}
\label{lumfr1}
\end{figure}
\noindent
{\bf FR~I:} Optical nuclear studies of radio  galaxies provide us with
crucial information on the physical processes at work in their central
regions.  Furthermore, a direct test  of the unification scheme, based
on the nuclear emission, can be performed. HST has allowed us to study
faint unresolved sources that are present in the  center of most FR~Is
from  the      3CR   catalog\cite{pap1}    (but   see    also  other
work\cite{capettib2,verdoes}).    We found that  these Central Compact
Cores   (CCC) show a tight   correlation with the  radio core emission
(both  in flux  and luminosity) which   strongly argues  for  a common
synchrotron  origin from both  components  (Fig.   \ref{lumfr1}).  The
detection of  CCCs in  85\% of  the complete  sample indicates that we
have a  direct view of the innermost  nuclear regions of  in the vast
majority  of FR~I.  Thus  it appears that a ``standard'' geometrically
and   optically thick torus is  not  present in   low luminosity radio
galaxies.   Any  absorbing  material     must be  distributed   in   a
geometrically thin structure (thickness over radius ratio $\lta 0.15$)
or, alternatively, thick tori are present only in  a minority of FR~I.
The CCC fluxes  are upper limits   to any thermal  disk emission.
This  implies extremely  low  radiative efficiency  for  the accretion
process.    The   observed  CCC   emission    corresponds   to   $\lta
10^{-7}$--$10^{-4}$   of  the   Eddington  luminosity    for  a $10^{8}
M_{\odot}$  black hole, which  appears to  be  typical for these radio
galaxies.  This  might also  explain  the lack of  strong photo-ionized
emission lines in the optical spectra of  both FR~Is and BL~Lacs.  The
picture which emerges  is that the  innermost structure of FR~I  radio
galaxies (and  thus also of  BL~Lacs) differs in many  crucial aspects
from that of the other classes of AGN; they  lack the substantial BLR,
tori and thermal disk emission, which  are instead associated with all
other active nuclei.

A fine  test for such a  scenario has been  obtained  in
the  case of   M~87,  a  famous  and  relatively powerful  FR~I  radio
galaxy\cite{perlman,whysong}. M~87 was observed in the infrared at $10
\mu$m with the Keck and  Gemini telescope, and no significant  mid--IR
nuclear  excess  is found.   This  supports   the absence of   a hidden
quasar-like accretion  process in  the  nucleus, which  would heat the
surrounding obscuring dust, thus producing  a strong thermal IR excess,
when compared  to the optical observed  flux. Of course  it would be
very interesting to extend this analysis to complete samples.

In  two cases (Centaurus~A and NGC~6251)   the nuclear spectral energy
distribution (SED)  has been derived from the  radio  to the gamma-ray
band.  The SEDs  show two broad  peaks, very similar to those observed
in BL~Lacs  and  usually  interpreted as  non-thermal  synchrotron and
inverse Compton emission\cite{ghisellini98}.  The SED has been modeled
in      the      framework      of    synchrotron         self-Compton
emission\cite{cena,6251,guainazzi},  obtaining physical parameters for
the source  that are completely in agreement  with those of BL~Lacs of
similar total    power,  thus  {\it   quantitatively  supporting   the
unification  scenario}.  Furthermore,  we  found  evidence   that  the
emitting region in FR~I has a slower bulk Lorentz factor ($\Gamma \sim
2$), when compared to BL~Lacs.  This can be interpreted as a signature
of the presence of velocity structures in the  jet, e.g.  a fast spine
and   a  slower (but  still   relativistic!)   layer similar to  those
observed in radio VLBI   images. The spine  dominates the  emission in
BL~Lacs, while the slower layer is visible only when the jet direction
forms a large angle to the observer's line-of-sight.

But    are   all FR~I   ``starved   quasars''?    In   fact, very  few
``broad-lined'' FR~Is, showing   the signature of  radiative efficient
accretion and substantial BLR, are known.  A famous example is 3C~120,
which is associated  to  a peculiar S0 galaxy\cite{tadhunter}.   A few
other FR~I quasars   have    been  recently  found in    deep    radio
samples\cite{blundell}. However it   is  still unclear  whether   they
represent a  substantial fraction of the   entire FR~I population that
have  been  ``hidden'' by a selection    bias, or they  are  only rare
peculiar sources.

\smallskip
\noindent
{\bf FR~II:} On  average, FR~II host  galaxies are less luminous  with
respect to those of FR~Is\cite{ledlowowen} (but  this is probably only
because of selection   effects\cite{scarpaurry}) and belong  to lower
density groups, at least at low redshifts\cite{zirbel97}.
One of the major unsolved issues for unification is the role played by
a sub-class of low-ionization FR~II (LEG).  Such objects have an FR~II
radio morphology, but their optical spectral properties are similar to
those of FR~I.   Our analysis  of  HST images of   the  nuclei of  3CR
FR~II\cite{pap4} up to $z=0.3$ led us to  intriguing conclusions.  The
first surprising result is that,  as in the  case of FR~I, most of the
galaxies  show central   unresolved  components,  regardless of  their
optical spectral classification.   In    the  framework of  the    AGN
unification scheme we would not  expect to observe such optical nuclei
in objects that do not show broad lines in their optical spectrum.  In
fact, their central regions should be hidden by  the presence of thick
obscuring  tori.   Broad-lined objects (quasars  and  broad line radio
galaxies,  BLO) have the brightest nuclei,  which show a large optical
excess  with respect to the radio-optical  correlation  found for FR~I
(Fig.  2a).   This is readily explained  if the  dominant component in
the optical  band is due to thermal  emission from  an accretion disk.
Indeed, these nuclei appear  to have flatter optical--to--UV  spectral
index ($\alpha_{o-UV}
\lta  1$)  when compared to  that  of ``synchrotron'' FR~I nuclei, and
similar  to that of  other radio--loud QSO\cite{papuv}.  We also found
that  optical nuclei  of  BLO are  present   only for  $L_{{\rm o}}  >
10^{28}$ erg s$^{-1}$ Hz$^{-1}$.  This might be the manifestation of a
threshold  in  the efficiency  of  the  accretion  process,  from  the
standard optically  thick,  geometrically thin accretion disk   to low
radiative accretion flows.

\begin{figure}[ht]

\epsfxsize=10.0cm   
\epsfbox{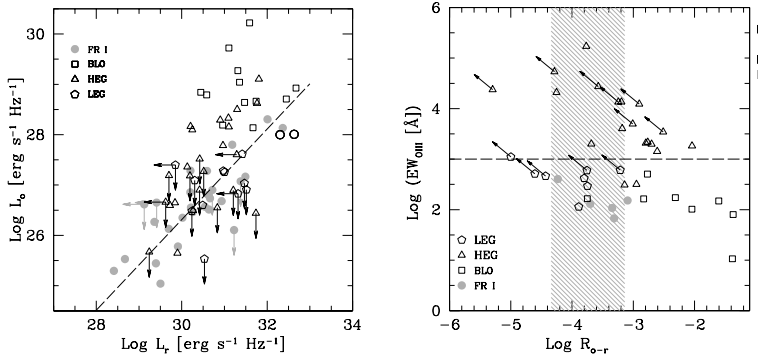}
%\hspace{1.5cm}
%\epsfxsize=4.5cm   
%\epsfbox{chiaberge_fig2b.ps}
\caption{{\bf Left:} (a) optical CCC vs. radio core luminosity for 3CR FR~I and FR~II with $z<0.3$. The dashed line is the FR~I correlation {\bf Right:} (b) Nuclear EW of the [OIII] emission line vs. the logarithm of the ratio between optical and radio core luminosity. The dashed lines separates scattered nuclei from nuclei seen directly.}
\end{figure}

The nature of  the  nuclei of the   High Excitation Galaxies (HEG)  is
certainly  more complex,  since some  of their nuclei   may be compact
scattering regions which fall  on the FR~I correlation ``by  chance''.
In  order to discriminate  between  scattered nuclei  and  nuclei seen
directly, we should  refer to an  isotropic parameter.  The luminosity
of the [OIII] emission line is believed to  be a good indicator of the
strength of the nuclear ionizing  continuum.  In Fig.  2b we show that
sources separate in   the plane formed by  the  ``nuclear'' EW of  the
[OIII] line vs the optical excess with respect  to the non-thermal jet
emission.  BLO, LEG and FR~I have low  EW values ($\sim 10^{2.5}$ \AA)
and they only  differ by the amount  of optical excess.  On the  other
hand, all but two of the HEG have much  larger equivalent widths (\gta
$10^{3.5}$ \AA).  This is indeed expected from  the unified models, as
obscuration reduces  the  observed nuclear continuum, while   the line
emission is less affected or  unaffected.  A strong ionization source,
obscured to our line-of-sight, must be  present in sources with a very
high value of  EW[OIII].  We argue  that all sources  with high EW are
hidden quasars, and their nuclei are seen through scattered radiation,
while the low  EW region of  the diagnostic plane includes the objects
in which we directly see the source of ionization.

It is important  to note that most  of the LEG  lie among  the FR~I in
both  diagnostic planes.  Therefore,  from the point  of view of their
radio-optical  nuclear properties, these objects are indistinguishable
from FR~I.  This picture leads to a  new dichotomy for radio galaxies,
which  is not based  on  their radio morphology  but  on their nuclear
properties.  The  LEGs should be considered  as part of the same group
as the FR~I.  A  direct implication for the  unifying scheme is  that,
when  observed along the   jet  axis, LEGs  should   appear as BL  Lac
objects.  Therefore, LEGs may be  identified as the parent  population
of  those BL~Lacs   with extended  radio morphology\cite{kollgaard92}.
Interestingly, both the  evolution and unification  of FR~I  and FR~II
with BL Lacs and  flat--spectrum quasars (FSRQ) can  be explained by a
dual--population scheme\cite{jacksonwall99}, which considers FR~Is and
LEGs as a single population, associated with  BL Lacs, while all other
FR~II  are unified to quasars.   And our HST results basically confirm
such a scenario.

\section{Is there any radio-loud AGN sequence?}
The properties of the  SED of blazars are well  described by  a single
parameter,   namely    the  bolometric   luminosity\cite{fossati}.  As
luminosity decreases, the frequency  of the synchrotron peak increases
as  well as the  ratio between  the luminosity  of the  Synchrotron to
inverse Compton peak.    This characterizes the ``blazars  sequence'',
from FSRQ to  low-energy-peaked  BL~Lacs  (LBL)  to high-energy-peaked
BL~Lacs (HBL), which have the synchrotron peak in  the X-rays.  Such a
trend appears  to  be  strictly connected  with  the  intensity of the
radiation  field surrounding the relativistic  jet\cite{ghisellini98}:
the position of the peaks of the SED is determined by the break of the
emitting particles  energy distribution, which is  located  at a lower
energy  for  higher  intensity   radiation  fields.  This  theoretical
scenario   implicitly  predicts that   high-energy-peaked blazars with
strong  emission lines   (HFSRQ)   should  not  exist.   Recently,   a
population   of  HFSRQ  might  have been  identified\cite{padovani03}.
However, although their peak energy appears to  be higher than that of
``classic'' FSRQ, $\nu_{peak}$  values are not  as extreme as those of
HBLs.   This means that there may  be a physical limit in $\nu_{peak}$
for sources with a strong radiation field surrounding the jet.  Can we
translate such a scenario into a physical sequence for radio galaxies?
It  is tempting to speculate  on the existence   of a similar sequence
starting from low-power FR~Is through HEGs  (and BLRGs), with the LEGs
acting as an intermediate class.  But it is unclear if the differences
between the various classes of RG are driven by e.g.   a change in the
physical conditions of the  accretion  process, which may  also affect
the properties of emission line field\cite{celottigg}.  I believe this
is one of  the most  interesting aspects of  the unifying  scheme  for
radio loud AGN to be investigated in years to come.

\vspace{-0.3cm}

\section*{Acknowledgments}

I apologize for the many  omissions of my talk:  the topic is immense,
and it  is impossible  to  be complete.  Most  of  the work  on  radio
galaxies' nuclei summarized here  have been done with the  fundamental
contribution  of other  people.    I   wish  to  thank  in  particular
Alessandro   Capetti,   Annalisa Celotti    and Duccio   Macchetto for
productive collaboration and constant  support.   Finally, I wish   to
congratulate the LOC and the SOC, and in  particular Raul and Roberto,
for organizing a very successful meeting in such a beautiful place.

\end{document}